\documentclass{article}

\usepackage{url}
\usepackage{amsmath}
\usepackage{amsthm,amsmath,amssymb}
\usepackage{bm}
\usepackage{graphicx}
\usepackage{float} 
\usepackage{subfigure}
\usepackage{algorithm}
\usepackage{algorithmic}
\usepackage{multirow}
\usepackage{orcidlink}
\usepackage{cite}
\usepackage{booktabs}
\usepackage{color}

\newcommand{\RR}{\mathbb{R}}
\newcommand{\PP}{\mathcal{P}}
\newcommand{\FF}{\mathcal{F}}
\newcommand{\xx}{\mathbf{x}}
\newcommand{\dv}{\mathbf{v}}
\newcommand{\uu}{\mathbf{u}}
\newcommand{\bb}{\mathbf{b}}
\newcommand{\etaa}{\bm{\eta}}

\newcommand{\te}{\bm{\theta}}
\newcommand{\G}{\mathbf{G}}
\newcommand{\ZZ}{\mathbf{Z}}
\newcommand{\WW}{\mathbf{W}}

\newcommand{\del}{\bm{\delta}}

\title{Untrained neural network embedded Fourier phase retrieval from few measurements}
\author{Liyuan Ma \thanks{
College of Science, National University of Defense Technology,
Changsha, Hunan, 410073, P.R.China. Email: \texttt{maliyuan21@nudt.edu.cn}}
\and Hongxia Wang {\thanks{
College of Science, National University of Defense Technology,
Changsha, Hunan, 410073, P.R.China. Corresponding author. Email: \texttt{wanghongxia@nudt.edu.cn}}}
\and Ningyi Leng {\thanks{
College of Science, National University of Defense Technology,
Changsha, Hunan, 410073, P.R.China. Email: \texttt{lengningyi14@nudt.edu.cn}}}
\and Ziyang Yuan {\thanks{Academy of Military Science of People’s Liberation Army, Beijing, P.R.China. Email: \texttt{yuanziyang11@nudt.edu.cn}}}
}

\date{}

\begin{document}
\maketitle


\begin{abstract}
    Fourier phase retrieval (FPR) is a challenging task widely used in various applications. It involves recovering an unknown signal from its Fourier phaseless measurements. FPR with few measurements is important for reducing time and hardware costs, but it suffers from serious ill-posedness. Recently, untrained neural networks have offered new approaches by introducing learned priors to alleviate the ill-posedness without requiring any external data. However, they may not be ideal for reconstructing fine details in images and can be computationally expensive. This paper proposes an untrained neural network (NN) embedded algorithm based on the alternating direction method of multipliers (ADMM) framework to solve FPR with few measurements. Specifically, we use a generative network to represent the image to be recovered, which confines the image to the space defined by the network structure. To improve the ability to represent high-frequency information, total variation (TV) regularization is imposed to facilitate the recovery of local structures in the image. Furthermore, to reduce the computational cost mainly caused by the parameter updates of the untrained NN, we develop an accelerated algorithm that adaptively trades off between explicit and implicit regularization. Experimental results indicate that the proposed algorithm outperforms existing untrained NN-based algorithms with fewer computational resources and even performs competitively against trained NN-based algorithms.
\end{abstract}

\section{Introduction}
In optics, most detectors can only record the magnitude or intensity of the signal, while losing the phase information \cite{Yuan2017}. Fourier phase retrieval (FPR) is an important inverse problem that seeks to recover an unknown signal $\xx^\ast \in \RR^n$ from its Fourier magnitude $\bb \in \RR^m$. The problem can be described as
\begin{equation} \label{problem}
    \mbox{Find } \xx \in \RR^n \mbox{ s.t. } \left| \FF \xx \right| + \del = \bb,
\end{equation}
where $\FF$ represents the Fourier transform operator, $\left| \cdot \right|$ denotes the element-wise absolute
value operator, and $\del$ is the additive noise. Note that a 2D image can also be expressed as $\xx^\ast \in \RR^n$ with $n_1\times n_2$ pixels, and the corresponding Fourier magnitude as $\bb \in \RR^m$ with $m_1\times m_2$ elements  by a lexicographical order, where $n_1\cdot n_2=n$ and $m_1\cdot m_2=m$. FPR arises in many applications, such as X-ray crystallography, ptychography, and diffraction imaging \cite{Jianwei2008, Guoan2021, Zhiming2023, Pham2018}.

FPR is an ill-posed inverse problem due to the non-uniqueness of its solution. Even though $m \ge 2n-1$, there can be up to $2^{n-2}$ solutions for Eq. (\ref{problem}), along with global phase shift, conjugate inversion, and spatial shift \cite{Beinert2015}. Most existing literature \cite{Wen2012, Metzler2018, wang2020deep, Mingqin2022} conducts simulation experiments under the oversampled condition of $m \ge 2n-1$. FPR with few measurements is seriously ill-posed, as the uniqueness of the solution lacks theoretical analysis under the condition of $m < 2n-1$. Classical algorithms such as Gerchberg–Saxton (GS) \cite{Gerchberg1972} and hybrid input and output (HIO) \cite{Fienup1982} often perform poorly under this condition. However, there are still many scenarios \cite{Maddali2019, Batey2014} where FPR with few measurements is common, especially when the detector's resolution is limited.

To overcome the ill-posedness of FPR, there are various regularization methods that introduce priors of the signal to FPR models. One of the common models is built as
\begin{equation} \label{reg}
    \hat{\xx} = \underset{\xx \in \RR^n}{\mathop{\arg\min}} \mbox{ } \frac{1}{2m} \Vert \bb - \left| \FF \xx \right| \Vert_2^2 + \alpha R(\xx),
\end{equation}
where the first and second terms are data fidelity and regularization terms respectively, and the regularization parameter $\alpha > 0$ pursues the tradeoff between these two terms. Explicit regularizers, such as $\delta_{\Omega}(\xx)$, $\Vert \xx \Vert_1$, and $\Vert \xx \Vert_{TV}$ \cite{Chen2007, Sarangi2017, Chang2018}, often capture certain properties of signals, such as support domain constraints, sparsity, and edge preservation. However, they are usually insufficient for representing complex information in images.

Recently, implicit regularizers based on learning have been proposed to represent more complicated priors of images. According to whether labeled data is required for training, this approach can be divided into supervised and unsupervised learning. \cite{Rakib2020} uses supervised learning with pairs of data $\left\{ \bb_i, \xx_i \right\}$ to train an end-to-end neural network (NN) that represents an implicit prior. The Plug-and-Play (PnP) framework \cite{Xue2022} leverages a pre-trained denoiser to implicitly represent a denoising prior. prDeep \cite{Metzler2018} adapts the Regularization by Denoising (RED) framework $R(\xx) = \frac{1}{2}\xx^T\left( \xx - D(\xx) \right)$ to solve phase retrieval (PR) problems. RED penalizes the residual difference between the image with its denoised self and the correlations between the image with the residual, where $D(\xx)$ denotes the pre-trained DnCNN \cite{Zhang2017} denoiser. In addition, trained generative networks \cite{Hand2018, Hyder2019, Shamshad2018} learn the distribution of images from labeled datasets to construct a generative prior. However, all of these supervised learning-based algorithms require acquiring a large amount of data with ground truth images. This can be expensive or even impossible in many fields. Moreover, limited generalization can easily occur when the distribution of the training and test data is inconsistent.

Unsupervised learning offers new approaches to impose implicit regularization without requiring labeled data. Among them, untrained NNs \cite{Adnan2023} propose a new paradigm for training randomly initialized NNs using only a single observation, without any external data. Untrained NNs, such as Deep Image Prior (DIP) \cite{Ulyanov2017} and its variant Deep Decoder (DD) \cite{Heckel2019}, demonstrate that their network architecture can capture statistical priors of an image from the observation itself. Net-PGD \cite{jagatapGauri2019} imposes an untrained generative prior by constraining $\xx$ to the generation space represented by a DD-based network, and solves Gaussian PR using the projected gradient descent (PGD) framework. It is beneficial for Gaussian PR with few measurements compared to hand-crafted priors. However, it is not ideal for solving more challenging FPR. DeepMMSE \cite{Mingqin2022} is an unsupervised learning-based algorithm which utilizes an untrained generative network with dropout to approximate the minimum mean squared error (MMSE) estimator of the image. Although it performs well in terms of FPR, its computational cost is quite high due to numerous iterations and a large number of network parameters. When $\xx$ is a $128\times 128$ grayscale image, $\bb$ has a size of $256\times 256$, and the network structures are set as specified in section \ref{expriment}, the number of network parameters in DeepMMSE is 708402, which is almost seven times that of Net-PGD, which has 108160 parameters. This makes it impractical to use when computational resources are limited. Therefore, it is necessary to explore unsupervised learning-based FPR algorithms that can achieve both fast and high-quality reconstruction.

\begin{figure}
    \centering
    \includegraphics[width = 12cm]{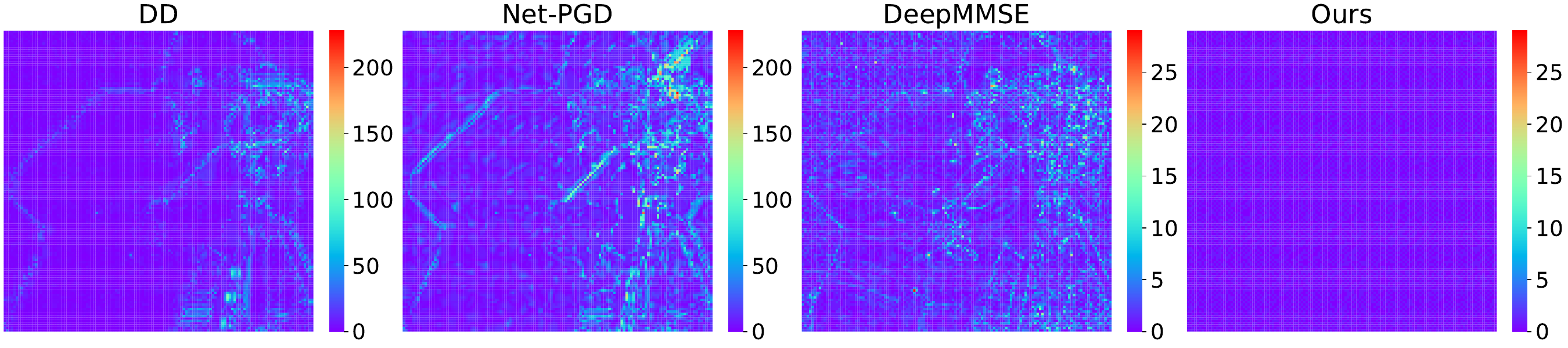}
    \caption{Comparison of different algorithms based on untrained NNs conducting FPR on \emph{Cameraman}. The networks do not require any external data for training. We present heat maps of reconstructed errors, which show the difference between the reconstructed image and the ground truth. The color scheme on the right side of the heat map represents the range of error values. The computational times for DD, Net-PGD, DeepMMSE and our algorithm are 58.48s, 68.23s, 739.99s, and 36.02s, respectively.}
    \label{fig:diff}
\end{figure}

To achieve fast and high-quality reconstruction of FPR using unsupervised learning, we adopt a DD-based network to impose a generative prior for FPR due to its latent ability to reconstruct the image from few measurements. However, unlike Net-PGD, we use the superior alternating direction method of multipliers (ADMM) \cite{Wen2012, LiSong2022} framework. As shown in Fig. \ref{fig:diff}, the DD algorithm, which employs a DD-based network to impose a generative prior for FPR within the ADMM framework, outperforms Net-PGD in the PGD framework. In addition, multiple studies \cite{Heckel2020, RHeckel2020} indicate that untrained NNs tend to fit smooth images and struggle to recover high-frequency structures. This is further supported by Fig. \ref{fig:diff}, which shows that three algorithms based on untrained NNs have large reconstruction errors on the high-frequency structures in the image. To address this problem, we combine the DD-based network with total variation (TV) regularization, which is beneficial for preserving the local structures of images \cite{Thomas2021, Jonathan2014}. Based on the ADMM framework, we propose a vanilla TV-regularized FPR algorithm with DD, which is called Vanilla-TRAD.

In addition, we develop an accelerated algorithm for achieving fast reconstruction of FPR. In Vanilla-TRAD, the generative network is trained to find a solution within its range for each iteration of FPR. This results in a high computational cost. We have observed that the later iterations of the generative network are unnecessary since they yield solutions with little improvement in quality. Therefore, we introduce an acceleration strategy inspired by the hybrid steepest descent (HSD) method \cite{RegevCohen2021, IsaoYamada2005}. This strategy dynamically adjusts whether to use DD in the iterations, which improves the reconstruction speed. The proposed algorithm is called Accelerated-TRAD. Fig. \ref{fig:diff} illustrates the benefits of our proposed Accelerated-TRAD in terms of reconstruction quality and computational time. The acceleration strategy offers a new approach to adaptively trading off between explicit and implicit regularization.

The performance of Vanilla-TRAD and Accelerated-TRAD is extensively evaluated under various settings. The experimental results of FPR with different measurement lengths and noise levels show that Accelerated-TRAD outperforms existing untrained NN-based algorithms with less computational cost and performs competitively against trained NN-based algorithms. Analysis of the parameters shows that Accelerated-TRAD is not sensitive to parameter selection in the acceleration strategy.

The rest of this paper is organized as follows. In section 2, Vanilla-TRAD and Accelerated-TRAD are introduced. In Section 3, we compare the proposed algorithms to state-of-the-art algorithms under various settings. We also analyze the selection of regularizers and parameters for the proposed algorithms. Section 4 is the conclusion. The code is available at \url{https://github.com/Liyuan-2000/TRAD.git}.

\section{The proposed methods}

\subsection{Vanilla TV-regularized FPR algorithm under untrained generative prior}

We consider a regularized FPR problem under untrained generative prior. According to Eq. (\ref{reg}), the problem can be described as
\begin{equation} \label{model}
    \begin{split}
        \underset{\xx, \te}{\mbox{min}} & \mbox{ } \frac{1}{2m} \Vert \bb - \left| \FF \xx \right| \Vert_2^2 + \alpha R(\xx),\\
        \mbox{s.t.} & \quad \G(\te) - \xx = 0,
    \end{split}
\end{equation}
where $R(\xx)$ is an explicit regularizer that enforces desired properties on $\xx$, while $\G(\te)$ is an untrained generative network with parameters $\te$ that restricts $\xx$ in $Range(\G)$.

\begin{figure}[htbp]
    \centering
    \includegraphics[width = 10cm]{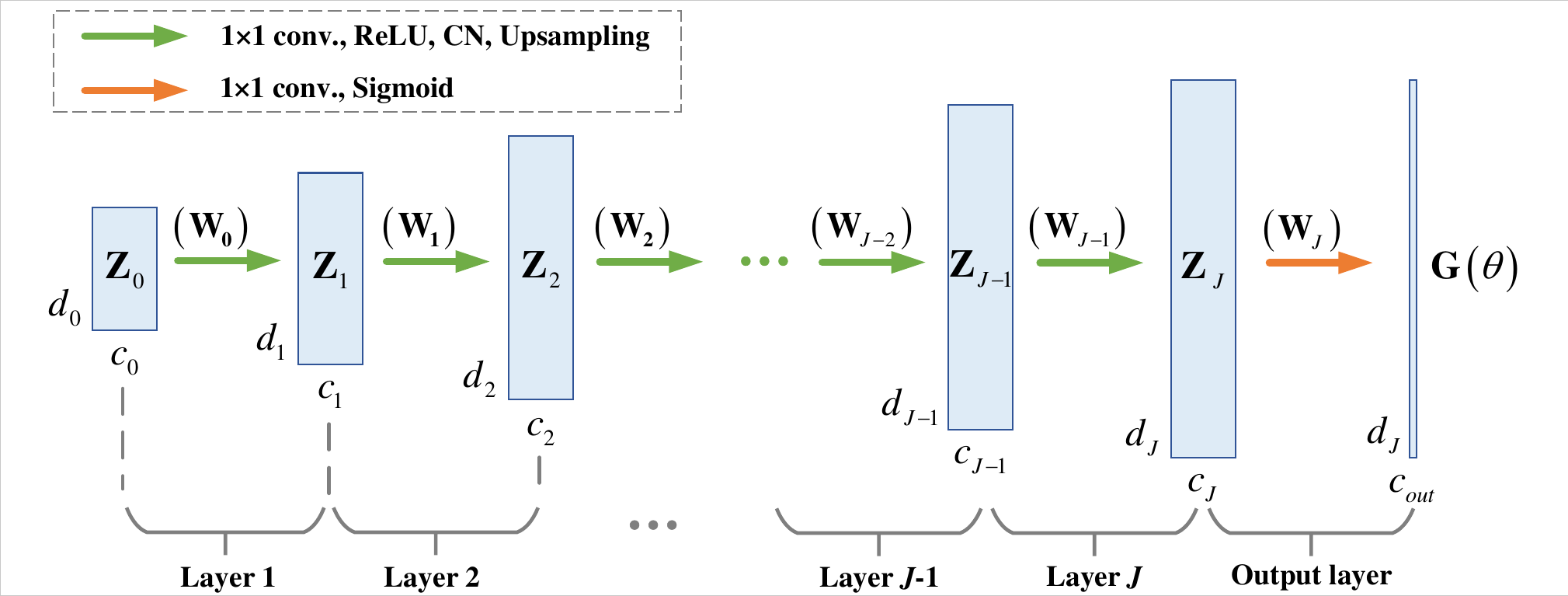}
    \caption{Illustration of the untrained generative network used in this paper. }
    \label{fig:network}
\end{figure}

An untrained generative network can implicitly capture priors of an image $\xx$ through its structure without external data \cite{Ulyanov2017, Heckel2019, Kevin2020}. In this paper, a fixed low-dimensional latent code $\ZZ_0 \in \RR^{d_0\times c_0}$ is chosen as the input of network. The output is $\G(\te) \in \RR^{d_J\times c_{out}}$, where $c_{out} = 1$ for a grayscale image, $c_{out} = 3$ for an RGB image, and $d_J \cdot c_{out}=n$. The structure of the network is illustrated by Fig. \ref{fig:network}. It is composed of $J$ layers, each of which consists of $1\times 1$ convolutions, a ReLU activation function $\mbox{relu}(\cdot)$, a channel normalization operator $\mbox{cn}(\cdot)$ and an upsampling operator $\mathbf{U}_j \in \RR^{d_{j+1}\times d_j}$. That is to say,
\begin{equation}
    \ZZ_{j+1} = \mathbf{U}_j \mbox{cn}\left( \mbox{relu} \left( \ZZ_j \WW_j \right) \right), j=0, 1, \cdots, J-1,
\end{equation}
where $\WW_j \in \RR^{c_{j}\times c_{j+1}}$ denotes the weight matrix corresponding to the $1\times 1$ convolutions.
The output layer of the network includes the $1\times 1$ convolutions and a sigmoid activation function $\mbox{sigmoid}(\cdot)$. It is expressed as:
\begin{equation}
    \G(\te) = \mbox{sigmoid}\left( \ZZ_J \WW_J \right),
\end{equation}
where $\WW_J \in \RR^{c_J \times c_{out}}$.
All parameters in the network can be vectored as $\te = \mbox{vec}\left( \WW_0, \WW_1, \cdots, \WW_J \right)$ and $\mbox{vec}(\cdot)$ is the vectorization of a matrix. For brevity, we utilize $\{ c_0, c_1, \cdots, c_J \}$ to represent the structure of a $J$-layer network. To reduce the cost of computation and storage, we adopt a network \cite{jagatapGauri2019} similar to DD \cite{Heckel2019}, where the sizes of the channels satisfy $d_0 < d_1 < \cdots < d_J$.

As mentioned before, while untrained generative networks are useful for reducing the ill-posedness of PR with few measurements, they struggle to recover high-frequency structures since they tend to fit smooth images \cite{Heckel2020}. We overcome this limitation by choosing a proper regularizer $R(\xx)$. For instance, $R(\xx) = \Vert \xx \Vert_{TV}$ encourages the preservation of local structures \cite{Thomas2021}. The TV norm $\Vert \xx \Vert_{TV}$ can be either anisotropic or isotropic, i.e., $\Vert \nabla \xx \Vert_1$ or $\Vert \nabla \xx \Vert_2$. Other regularization terms that protect high-frequency characteristics can also be considered for selection.

Eq. (\ref{model}) is a non-convex optimization problem with multiple local minima. This makes it challenging for general gradient-based algorithms to find the global optimal solution. To solve Eq. (\ref{model}), we use the ADMM \cite{Wen2012, Hong2016, Wang2019} framework, which mitigates the difficulty by dividing the problem into several subproblems that can be solved more easily. Note that the data fidelity term in Eq. (\ref{model}) is not Lipschitz differentiable, making it difficult to design efficient algorithms that use gradients. To address this, we replace it with its smoothed version \cite{Chang2019}:
\begin{equation} \label{f}
    f(\uu)=\frac{1}{2m}\Vert \sqrt{\bb^2+\varepsilon\mathbf{1}} - \sqrt{|\FF \uu|^2+\varepsilon \mathbf{1}} \Vert_2^2,
\end{equation}
where $\varepsilon > 0$ represents the penalization parameter, and $\mathbf{1} \in \RR^m$ denotes a vector whose elements are all ones. The gradient of $f(\uu)$ is
\begin{equation} \label{grad}
    \nabla f(\uu) = \uu - \FF^{-1}\left( \frac{\sqrt{\bb^2+\varepsilon \mathbf{1}}}{\sqrt{|\FF \uu|^2+\varepsilon \mathbf{1}}} \odot \FF \uu \right),
\end{equation}
where $\odot$ denotes the Hadamard product and $\FF^{-1}$ denotes the inverse Fourier transform operator.

Now we reformulate Eq. (\ref{model}) into
\begin{equation}\label{re-model}
    \begin{split}
        \underset{\xx, \te}{\mbox{min}} & \mbox{ } f\left(\G(\te)\right) + \alpha \Vert \xx \Vert_{TV},\\
        \mbox{s.t.} & \quad \G(\te) - \PP \xx=0,
    \end{split}
\end{equation}
where $\PP:\RR^n \to \RR^m$ is a zero-padding operator, which appends $m-n$ zeros after the last element of $\xx$. Introducing $\PP$ aims to execute the Fourier transform using the Fast Fourier transform (FFT) algorithm.

By means of the augmented Lagrangian multiplier method, we transform Eq. (\ref{re-model}) into an unconstrained optimization
problem:
\begin{equation}
    \underset{\etaa}{\mathop{\max}}\underset{\te, \xx}{\mathop{\min}} \mbox{ } L_{\rho}(\te,\xx,\etaa) = f(\G(\te))+\alpha \Vert \xx \Vert_{TV} + \langle \etaa, \G(\te)-\PP \xx \rangle +\frac{\rho}{2}\Vert \G(\te)-\PP \xx \Vert_2^2,
\end{equation}
where $\etaa \in \RR^m$ is the Lagrangian multiplier associated with the equality constraint, and $\rho > 0$ is the coefficient of the quadratic penalty term. Following the ADMM framework, we can minimize $L_{\rho}(\te,\xx,\etaa)$ with respect to $\te$, $\xx$, and $\etaa$ in an alternating way. Given initial variables $\xx_0$, $\etaa_0$ and $k=0$,
\begin{subequations}
    \begin{align}
        \te_{k+1} & = \underset{\te }{\mathop{\arg\min}} \mbox{ } \left\{ f(\G(\te)) + \frac{\rho}{2}\Vert \G(\te)-\PP \xx_k + \frac{\etaa_k}{\rho} \Vert_2^2 \right\} , \label{te}\\
        \xx_{k+1} & = \underset{\xx }{\mathop{\arg\min}} \mbox{ } \left\{ \alpha \Vert \xx \Vert_{TV} +\frac{\rho}{2}\Vert \G(\te_{k+1})-\PP \xx + \frac{\etaa_k}{\rho} \Vert_2^2 \right\}, \label{x}\\
        \etaa_{k+1} & = \etaa_k + \rho \left( \G(\te_{k+1}) - \PP \xx_{k+1} \right). \label{eta}
    \end{align}
\end{subequations}

The first subproblem (\ref{te}) is used to update the network parameters. We solve it in two steps. First, we calculate a vector $\uu \in \RR^m$, which acts a substitute for $\G(\te)$ in Eq. (\ref{te}). Since $f$ is differentiable, we make a first-order approximation of $f$ at $\PP \xx_k - \frac{\etaa_k}{\rho}$ and then calculate $\uu_{k+1}$ using the Karush-Kuhn-Tucker (KKT) condition:
\begin{equation} \label{u}
    \begin{split}
        \uu_{k+1} & = \underset{\uu }{\mathop{\arg\min}} \left\{ f(\uu) + \frac{\rho}{2}\Vert \uu-\PP \xx^k + \frac{\etaa_k}{\rho} \Vert^2 \right\} \\
        \quad & = \underset{\uu }{\mathop{\arg\min}} \mbox{ }  f\left(\PP \xx_k - \frac{\etaa_k}{\rho}\right) + \left \langle \nabla f\left( \PP \xx_k - \frac{\etaa_k}{\rho} \right), \uu - \PP \xx_k + \frac{\etaa_k}{\rho} \right \rangle \\
        \quad & + \frac{\rho}{2}\Vert \uu -\PP \xx_k + \frac{\etaa_k}{\rho} \Vert^2 \\
        \quad & = \PP \xx_k - \frac{\etaa_k}{\rho} - \frac{1}{\rho} \nabla f\left(\PP \xx_k - \frac{\etaa_k}{\rho}\right).
    \end{split}
\end{equation}
The second step is to project $\uu_{k+1}$ into $Range(\G)$ by searching for a suitable $\te_{k+1}$, which serves as an internal loop at the $k+1$-th iteration.
\begin{equation} \label{tt}
    \te_{k+1} = \underset{\te}{\mathop{\arg\min}} \mbox{ } \Vert \G(\te) - \uu_{k+1} \Vert_2^2.
\end{equation}
The detailed process for solving Eq. (\ref{tt}) will be explained in subsection \ref{B}. For convenience, we denote $\dv_{k+1} = \G\left(\te_{k+1}\right)$ and substitute $\dv_{k+1}$ for $\G\left(\te_{k+1}\right)$ in Eq. (\ref{x}) and Eq. (\ref{eta}).

Now consider the second subproblem (\ref{x}). We make a first-order approximation of $\Vert \cdot \Vert_{TV}$ at $\PP^{-1} \left( \dv_{k+1} + \frac{\etaa_k}{\rho} \right)$ again by the gradient $ \nabla \Vert \cdot \Vert_{TV}$ and then calculate $\xx_{k+1}$ using the KKT condition. The result is
\begin{equation} \label{xx}
    \xx_{k+1} = \PP^{-1} \left( \dv_{k+1} + \frac{\etaa_k}{\rho} \right) - \frac{\alpha}{\rho} \Delta \left[ \PP^{-1}  \left( \dv_{k+1} + \frac{\etaa_k}{\rho} \right) \right],
\end{equation}
where $\Delta$ denotes the Laplace operator due to $\nabla \Vert \xx \Vert_{TV} = \mbox{div} \left( \frac{\nabla \xx}{ \Vert \xx \Vert} \right)  = \frac{1}{\Vert \xx \Vert} \Delta \xx $ and $\mbox{div}(\cdot)$ denotes the divergence.

As to the initialization, we just set $\xx_0 = \FF^{-1} ( \bb )$ and $\etaa_0 = \mathbf{0}$. Thus, we have developed an algorithm to solve FPR, called the Vanilla TV-regularized FPR algorithm with DD (Vanilla-TRAD). The detailed steps are summarized in Algorithm \ref{alg1}.

\begin{algorithm}[htbp]
    \renewcommand{\algorithmicrequire}{\textbf{Input:}}
	\renewcommand\algorithmicensure {\textbf{Output:} }
    \caption{Vanilla-TRAD}
    \label{alg1}
    \begin{algorithmic}
        \REQUIRE $\bb, \xx_0 = \FF^{-1} ( \bb ), \etaa_0 = \mathbf{0}, \rho>0, \varepsilon>0$, $K$, $\te_0$\\
        \ENSURE $\xx_K$ \\
    \end{algorithmic}
    \vskip 2mm
    \hrule
    \vskip 2mm
    \begin{algorithmic}[1]
        \STATE $\mathbf{for}\quad k = 0$ : $K-1$ \\
        \STATE \quad \quad $\uu_{k+1} = \PP \xx_k - \frac{\etaa_k}{\rho} - \frac{1}{\rho} \nabla f\left(\PP \xx_k - \frac{\etaa_k}{\rho}\right)$ \\ 
        \STATE \quad \quad $\te_{k+1} = \underset{\te}{\mathop{\arg\min}} \mbox{ } \Vert \G(\te) - \uu_{k+1} \Vert_2^2$ \\
        \STATE \quad \quad $\dv_{k+1} = \G\left(\te_{k+1}\right)$ \\
        \STATE \quad \quad $\xx_{k+1} = \PP^{-1} \left( \dv_{k+1} + \frac{\etaa_k}{\rho} \right) - \frac{\alpha}{\rho} \Delta \left[ \PP^{-1}  \left( \dv_{k+1} + \frac{\etaa_k}{\rho} \right) \right]$ \\
        \STATE \quad \quad $\etaa_{k+1} = \etaa_k + \rho \left( \dv_{k+1} - \PP \xx_{k+1} \right)$ \\
        \STATE $\mathbf{end}$
    \end{algorithmic}
\end{algorithm}

\subsection{Accelerated TV-regularized FPR algorithm under untrained generative prior} \label{B}

Vanilla-TRAD effectively combines TV regularization and the untrained generative prior through the ADMM framework. However, its computational cost may be expensive because solving Eq. (\ref{tt}) is required to update the network parameters for each $k$, even though parameter reduction has already been considered in the selection of network structure. In this subsection, we propose an acceleration algorithm to improve the efficiency of Vanilla-TRAD.

Before introducing the acceleration algorithm, we first discuss the specific implementation of Eq. (\ref{tt}) in Vanilla-TRAD. The loss function is $g(\te) = \Vert \G(\te) - \uu_{k+1} \Vert_2^2$. Given the initial network parameters $\te_0$, learning rate $\gamma_0$, and the number of internal loops $l_0$, the parameters can be updated by minimizing $g(\te)$ using gradient descent,
\begin{equation} \label{para}
    \te_{k+1}^{(j+1)} = \te_{k+1}^{(j)} - \gamma_k \nabla g\left( \te_{k+1}^{(j)} \right), j = 0, 1, \cdots, l_k-1,
\end{equation}
where $\te_{k+1}^{(0)} = \te_k$ and $\te_{k+1} = \te_{k+1}^{(l_k)}$. Variants of gradient descent, such as SGD \cite{bottou1991} and Adam \cite{konur2015}, can also be used. The learning rate $\gamma_k$ decreases as $k$ increases, following the formula $\gamma_k = \gamma_0 \beta^{\lfloor k / \kappa_1 \rfloor}$, where $0<\beta<1$ denotes the decay factor and $\kappa_1$ represents the epoch of decay. The number of internal loops, denoted by $l_k$, is calculated as $l_k = \mbox{round} \left( l_0 \zeta^{ \lfloor k/\kappa_2 \rfloor} \right)$, where $\zeta>1$ denotes the growth factor, $\kappa_2$ represents the epoch of growth, and $\mbox{round}(\cdot)$ is the integer-valued function. Note that the number of internal loops $l_k$ gradually increases with $k$, so the computational time for each iteration also increases with $k$.

Referring to the HSD method \cite{IsaoYamada2005}, we modify the 4th line of Vanilla-TRAD to a weighted combination of $\uu_{k+1}$ and $\G(\te_{k+1})$.
\begin{equation} \label{weight}
    \dv_{k+1} = \mu_k \G\left(\te_{k+1}\right) + (1-\mu_k)\uu_{k+1}.
\end{equation}

\begin{figure}
    \centering
    \includegraphics[width=7cm]{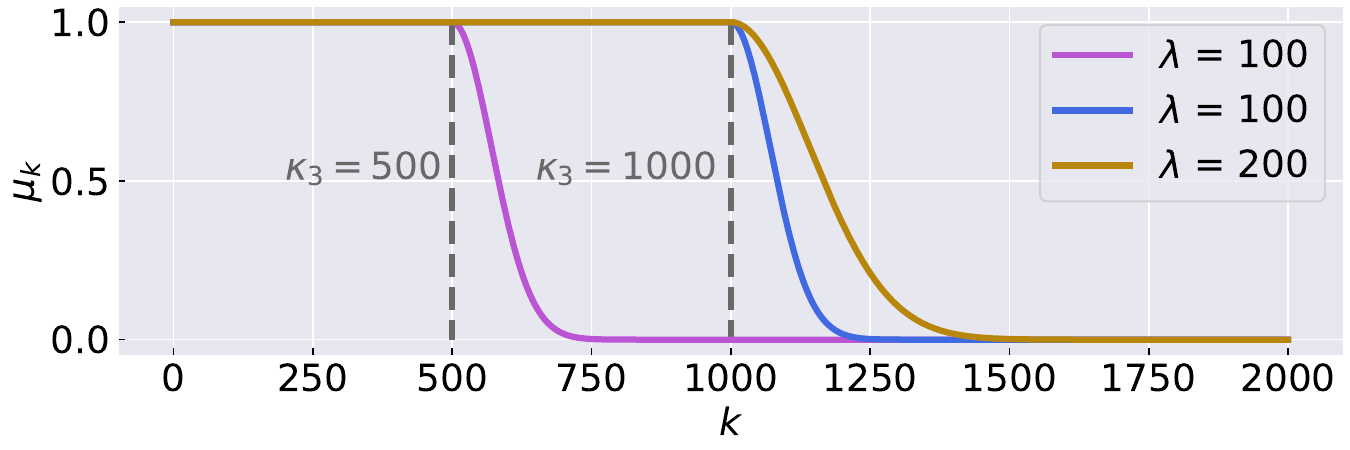}
    \caption{The change of weight $\mu_k$ with different $\lambda$ and $\kappa_3$.}
    \label{fig:mu}
\end{figure}

During the early iterations, we expect to play the role of the untrained generative prior to obtain a relatively stable intermediate solution. Once stability is reached, the weight $\mu_k$ can be reduced to 0, since it is unnecessary to project $\uu_{k+1}$ into $Range(\G)$ when it is close to the optimal solution. Thus, the weight $\mu_k$ can be designed to be 1 in the early iterations and decrease to 0 later. We choose a specific formula given by
\begin{equation} \label{mu}
    \mu_k = \mbox{exp}\left\{ - \left( \frac{\mbox{max}\left\{0, k - \kappa_3 \right\}}{\lambda} \right)^2 \right\},
\end{equation}
where $\kappa_3$ denotes the iteration step at which the weight $\mu_k$ begins to decrease, and $\lambda$ represents the rate of decrease. As shown in Fig. \ref{fig:mu}, a small $\kappa_3$ indicates early weight decay, while a small $\lambda$ indicates fast weight decay.

\begin{algorithm}[htbp]
    \renewcommand{\algorithmicrequire}{\textbf{Input:}}
	\renewcommand\algorithmicensure {\textbf{Output:} }
    \caption{Accelerated-TRAD}
    \label{alg2}
    \begin{algorithmic}
        \REQUIRE $\bb, \xx_0 = \FF^{-1} ( \bb ), \etaa_0 = \mathbf{0}, \rho>0, \varepsilon>0$, $K$, $\te_0$, $\gamma_0$, $l_0$, $\beta$, $\zeta$, $\lambda$, $\kappa_1$, $\kappa_2$, $\kappa_3$\\
        \ENSURE $\xx_K$ \\
    \end{algorithmic}
    \vskip 2mm
    \hrule
    \vskip 2mm
    \begin{algorithmic}[1]
        \STATE $\mathbf{for}\quad k = 0$ : $K-1$ \\
        \STATE \quad \quad $\uu_{k+1} = \PP \xx_k - \frac{\etaa_k}{\rho} - \frac{1}{\rho} \nabla f\left(\PP \xx_k - \frac{\etaa_k}{\rho}\right)$ \\ 
        \STATE \quad \quad $\te_{k+1}^{(0)} = \te_k$ \\
        \STATE \quad \quad $\gamma_k = \gamma_0 \beta^{\lfloor k/\kappa_1 \rfloor}$ \\
        \STATE \quad \quad $l_k = \mbox{round} \left( l_0 \zeta^{ \lfloor k/\kappa_2 \rfloor} \right)$ \\
        \STATE \quad \quad $\mathbf{for}\quad j = 0$ : $l_k-1$ \\
        \STATE \quad \quad \quad \quad $\te_{k+1}^{(j+1)} = \te_{k+1}^{(j)} - \gamma_k \nabla g\left( \te_{k+1}^{(j)} \right)$ \\
        \STATE \quad \quad $\mathbf{end}$ \\
        \STATE \quad \quad $\te_{k+1} = \te_{k+1}^{(l_k)}$ \\
        \STATE \quad \quad $\mu_k = \mbox{exp}\left\{ - \left( \frac{\mbox{max}\left\{0, k - \kappa_3 \right\}}{\lambda} \right)^2 \right\}$ \\
        \STATE \quad \quad $\dv_{k+1} = \mu_k \G\left(\te_{k+1}\right) + (1-\mu_k)\uu_{k+1}$ \\
        \STATE \quad \quad $\xx_{k+1} = \PP^{-1} \left( \dv_{k+1} + \frac{\etaa_k}{\rho} \right) - \frac{\alpha}{\rho} \Delta \left[ \PP^{-1} \left( \dv_{k+1} + \frac{\etaa_k}{\rho} \right) \right]$ \\
        \STATE \quad \quad $\etaa_{k+1} = \etaa_k + \rho \left( \dv_{k+1} - \PP \xx_{k+1} \right)$ \\
        \STATE $\mathbf{end}$
    \end{algorithmic}
\end{algorithm}

Note that Accelerated-TRAD removes the network in later iterations. This benefits both the preservation of high-frequency structures in images and the reduction of computational cost. The proposed accelerated algorithm, referred to as Accelerated-TRAD, is shown in Algorithm \ref{alg2}.

\textbf{Discussions.}
1. Accelerated-TRAD is an accelerated version of Vanilla-TRAD. For the first $\kappa_3$ iterations, the format of Accelerated-TRAD is consistent with that of Vanilla-TRAD. In the case of $\mu_k \equiv 1$, Accelerated-TRAD is equivalent to Vanilla-TRAD.

2. The main complexity of the proposed algorithms lies in the internal loops implemented by $\G(\te)$. In Accelerated-TRAD, the removal of the network in some iterations can effectively reduce computational complexity. Additionally, removing the network in later iterations is more effective than doing so earlier. This is because there are more internal loops in the later iterations, resulting in a higher time cost.

3. It seems that Accelerated-TRAD has added two parameters compared to Vanilla-TRAD. In fact, they can be simply chosen as $\kappa_3=1000$ and $\lambda=10$. It will be demonstrated through experiments in subsection \ref{impact} that the algorithm is not sensitive to parameter selection.

\section{Experiment}\label{expriment}
Performance evaluation is conducted on FPR at different measurement lengths and noise levels. We compare the proposed algorithms with state-of-the-art algorithms on PR, especially learning-based methods, including prDeep \cite{Metzler2018}, Net-PGD \cite{jagatapGauri2019} and DeepMMSE \cite{Mingqin2022}.

\begin{itemize}
    \item prDeep \cite{Metzler2018}: A supervised learning algorithm based on the pre-trained DnCNN \cite{Zhang2017} is used to solve PR problems with different measurement models and noise levels. The algorithm integrates the DnCNN denoiser into the RED framework with the HIO initialization.
    \item Net-PGD \cite{jagatapGauri2019}: An unsupervised learning-based algorithm combines the projected gradient descent (PGD) algorithm with a DD-based network to solve Gaussian PR. The algorithm implements gradient descent using external loops to produce the estimated solution, which is then projected into the network space using internal loops.
    \item DeepMMSE \cite{Mingqin2022}: An unsupervised learning-based algorithm utilizes an untrained generative network with dropout to approximate the minimum mean squared error (MMSE) estimator of the image in PR. The algorithm also requires the HIO initialization, as described in \cite{Metzler2018}.
\end{itemize}

\begin{figure}[htbp]
    \centering
    \includegraphics[width = 6cm]{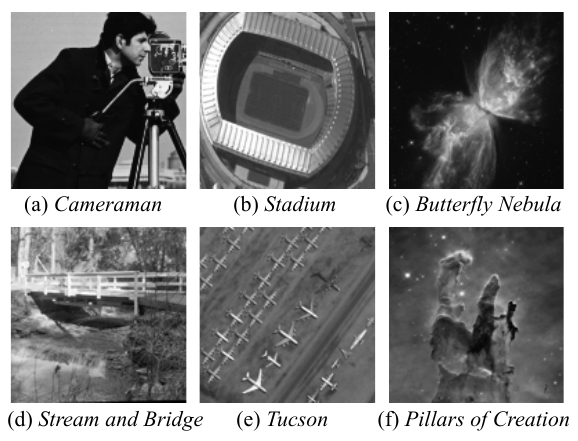}
    \caption{The data used in the experiment. First column: 2 natural images. Second column: 2 remote sensing images. Third column: 2 microscopic images.}
    \label{fig:images}
\end{figure}

The experimental data, shown in Fig. \ref{fig:images}, consists of $128\times 128$ grayscale images cropped or resized from standard $256\times 256$ images, including 2 natural images, 2 remote sensing images, and 2 microscopic images. We evaluate the performance of the compared algorithms using the Peak Signal to Noise Ratio (PSNR), Structural Similarity (SSIM) of the reconstructed results, as well as the time cost of calculation. Due to the randomization factors involved in initializing and updating network parameters, the algorithms may produce different results in different runs. To ensure a fair comparison, each algorithm is run five times. The average PSNR, SSIM and time cost of the five reconstructed results are then calculated.

The parameters in the proposed algorithms are as follows. We use isotropic TV norm in our experiments. The initial learning rate is $\gamma_0 = 0.005$, and the decay factor and epoch are $\beta = 0.5$ and $\kappa_1 = 500$. The initial number of internal loops is $l_0 = 5$, and the growth factor and epoch are $\zeta = 1.2$ and $\kappa_2 = 500$. The parameters $\te_0$ in the network are initialized according to He initialization \cite{He2015}. The elements of the latent code $\ZZ_0$ are generated from the Gaussian distribution $\mathcal{N}(0, 0.01)$. Additionally, we set $\rho = 1$, $\epsilon = 0.001$ and $\alpha = \frac{1}{384}$ by default. The structure of the network is designed to $\{128, 128, 128, 128\}$, which is a 3-layer network. The parameters in Accelerated-TRAD are set to $\kappa_3 = 1000$ and $\lambda = 10$, which will be analyzed in subsection \ref{impact}.

Following \cite{Mingqin2022}, the network input of DeepMMSE are generated from the Gaussian distribution $\mathcal{N}(0, 0.1)$. For Net-PGD, its network structure and internal loops are consistent with our algorithms. The learning rate of its external loops is initialized to 0.5, and the decay factor and epoch are set to 0.7 and 500, respectively. $\sigma_w$ in prDeep is set to 0.1 when no noise is added to the Fourier magnitude. Both Net-PGD and the proposed algorithms are conducted 2000 iterations while DeepMMSE is conducted 50000 iterations. prDeep is executed for 200 iterations four times, once for each of the denoiser networks trained at standard deviations of 50, 40, 20, and 10. All NNs adopt the Adam optimizer. All algorithms are conducted on NVIDIA A-100 GPUs. Among them, prDeep uses MATLAB R2021a, while the other algorithms use the Pytorch framework with Python 3.9.

\subsection{FPR at different sampling ratios}

\begin{figure}
    \centering
    \includegraphics[width = 13cm]{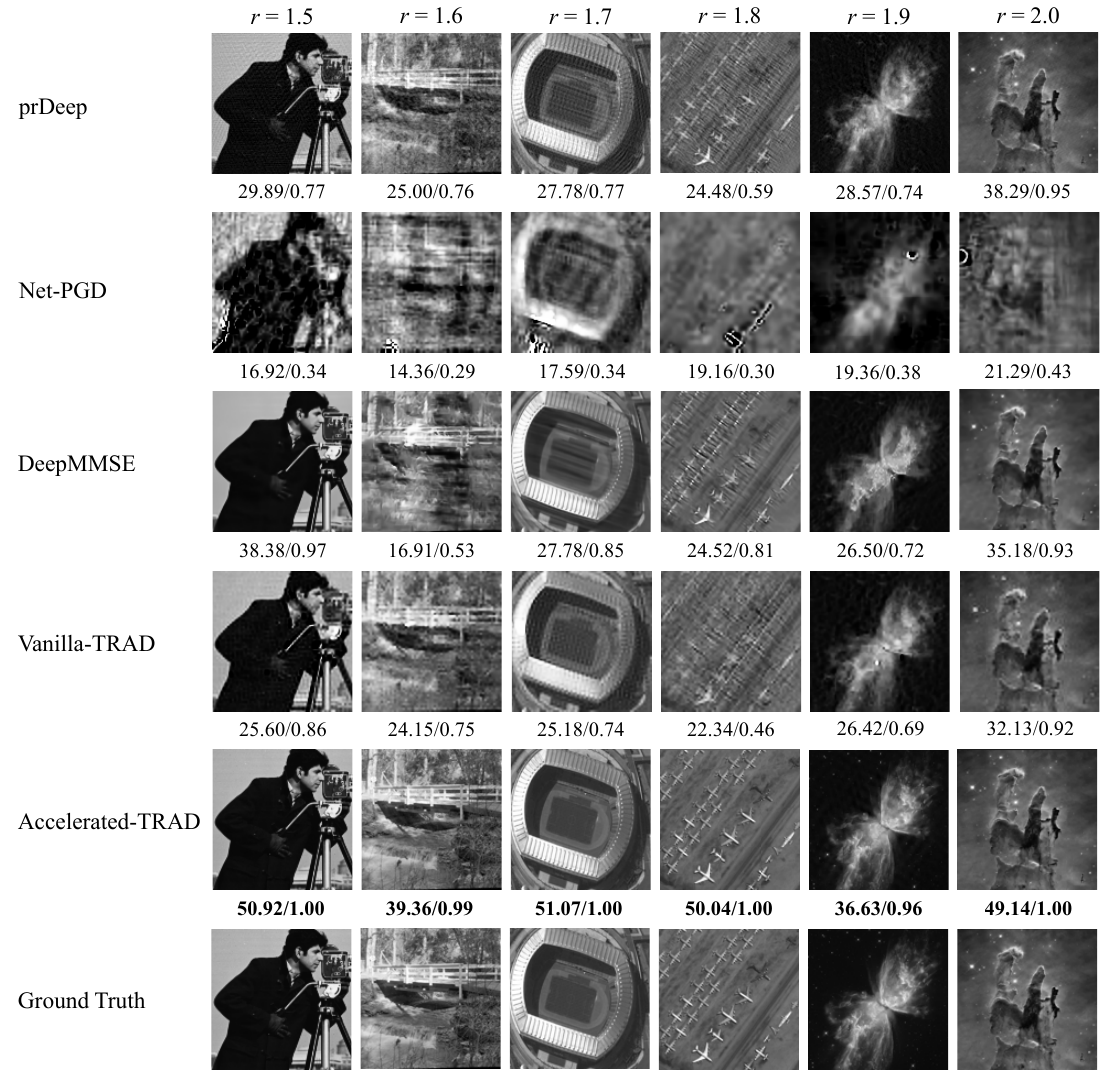}
    \caption{The reconstructed results of different algorithms at various sampling ratios, along with the corresponding PSNR(dB)/SSIM for each image annotated below.}
    \label{fig:compare_fig}
\end{figure}

This subsection compares algorithms for FPR at six sampling ratios ranging from 1.5: 0.1: 2.0. The sampling ratio, denoted as $r$, represents the ratio of the measurement length to image length in each dimension. For 2D images with equal width and height, $r=\sqrt{m}/\sqrt{n}$. Table \ref{tab1} shows the quantitative results for the compared algorithms at sampling ratios of 1.7 and 1.9. Additionally, a visual comparison between the algorithms at six sampling ratios can be seen in Fig. \ref{fig:compare_fig}. For each algorithm, the images with the highest PSNR in five runs are shown.

\begin{table}[htbp]
    \centering
    \caption{The PSNR, SSIM and computational time of the reconstructed results by different algorithms at sampling ratios of 1.7 and 1.9. The best and second best results at each column are \textbf{boldfaced} and \underline{underlined} respectively.}
    \label{tab1}
    \resizebox{12cm}{2.3cm}{
    \begin{tabular}{lccccccc}
        \toprule
        \multirow{2}{*}{$r$=1.7} & \multicolumn{6}{c}{PSNR(dB)/SSIM} & Time (s) \\
          & Cameraman & Stream and Bridge & Stadium & Tucson & Butterfly Nebula & Pillars of Creation & Total\\
        \midrule
        prDeep \cite{Metzler2018} & \underline{40.30}/0.95 & \underline{35.64/0.88} & 24.26/0.64 & 21.35/0.34 & \textbf{22.62/0.52} & \underline{25.62/0.66} & \textbf{29.78} \\
        Net-PGD \cite{jagatapGauri2019} & 14.87/0.28 & 13.24/0.25 & 13.42/0.22 & 19.00/0.28 & 19.86/0.38 & 18.71/0.31 & 57.21 \\
        DeepMMSE \cite{Mingqin2022} & 38.53/\underline{0.97} & 20.24/0.55 & \underline{25.22/0.75} & \underline{24.57}/\textbf{0.59} & 19.34/0.42 & 20.86/0.48 & 784.86 \\
        Vanilla-TRAD & 25.92/0.89 & 22.66/0.70 & 21.76/0.63 & 21.53/0.37 & \underline{21.59/0.48} & 24.58/0.61 & 65.24 \\
        Accelerated-TRAD & \textbf{51.12/1.00} & \textbf{45.50/1.00} & \textbf{31.03/0.79} & \textbf{25.09}/\underline{0.57} & 19.93/0.45 & \textbf{37.67/0.92} & \underline{33.93} \\
        \midrule
        \multirow{2}{*}{$r$=1.9} & \multicolumn{6}{c}{PSNR(dB)/SSIM} & Time (s) \\
         & Cameraman & Stream and Bridge & Stadium & Tucson & Butterfly Nebula & Pillars of Creation & Total\\
        \midrule
        prDeep \cite{Metzler2018} & \underline{50.33/0.98} & \underline{39.43/0.91} & \underline{34.54/0.91} & \underline{25.84/0.64} & 23.98/0.59 & 29.04/0.78 & \textbf{26.63} \\
        Net-PGD \cite{jagatapGauri2019} & 16.58/0.35 & 13.00/0.24 & 12.17/0.19 & 19.25/0.27 & 18.61/0.34 & 19.15/0.33 & 56.53 \\
        DeepMMSE \cite{Mingqin2022} & 38.58/0.97 & 24.73/0.77 & 31.63/0.90 & 24.20/0.58 & 23.66/0.60 & 29.83/0.82 & 775.82 \\
        Vanilla-TRAD & 26.08/0.90 & 25.69/0.82 & 25.28/0.76 & 22.38/0.49 & \underline{24.41/0.61} & \underline{30.35/0.86} & 65.30 \\
        Accelerated-TRAD & \textbf{51.14/1.00} & \textbf{49.26/1.00} & \textbf{51.13/1.00} & \textbf{33.53/0.68} & \textbf{29.11/0.82} & \textbf{46.39/0.99} & \underline{35.04} \\
        \bottomrule
    \end{tabular}}
\end{table}

Table \ref{tab1} and Fig. \ref{fig:compare_fig} indicate that the reconstructed results of Accelerated-TRAD have almost the highest PSNR, SSIM and visual quality. Among all unsupervised learning-based algorithms, Accelerated-TRAD performs the best in terms of both the quality of the reconstructed results and computational time. It also performs competitively against supervised learning-based prDeep with only a minimal increase in computational time, no more than 10 seconds.

\subsection{FPR from noisy measurements}
This subsection compares the robustness of algorithms for FPR. The measurement model
with noise is:
\begin{equation}
    \bb = \left| \FF \xx \right| + \del,
\end{equation}
where
$\del \sim \mathcal{N}(0, \sigma^2)$ stands for Gaussian noise with a standard deviation of $\sigma$.

When the sampling ratios are fixed, we compare algorithms at different noise levels. The noise level is represented by the Signal Noise Ratio (SNR), which is defined as
\begin{equation} \label{SNR}
    \mbox{SNR} = 20\mbox{log}_{10}\frac{\mbox{Var}\left( \left| \FF \xx \right| \right)}{\sigma^2}, 
\end{equation}
where $\mbox{Var}\left(\left| \FF \xx \right|\right)$ is the variance of $\left| \FF \xx \right|$. A low SNR indicates a high level of noise.

\begin{table}
    \centering
    \caption{The PSNR and SSIM of the reconstructed results by different algorithms on \emph{Pillars of Creation} at different noise levels. The best and second best results at each column are \textbf{boldfaced} and \underline{underlined} respectively.}
    \label{tab:tab2}
    \resizebox{12cm}{1.4cm}{
    \begin{tabular}{lcccccc}
        \toprule
         & \multicolumn{3}{c}{PSNR(dB)/SSIM ($r$=1.6)} & \multicolumn{3}{c}{PSNR(dB)/SSIM ($r$=1.8)} \\
         & SNR=30 & SNR=20 & SNR=10 & SNR=30 & SNR=20 & SNR=10 \\
        \midrule
        prDeep \cite{Metzler2018} & 19.34/0.30 & 18.80/\underline{0.29} & 18.00/0.22 & 18.88/0.30 & 19.57/0.31 & 18.75/0.25 \\
        Net-PGD \cite{jagatapGauri2019} & 18.94/0.31 & 18.41/0.23 & 17.24/0.15 & 19.17/0.35 & 18.70/0.28 & 18.14/0.18 \\
        DeepMMSE \cite{Mingqin2022} & 19.29/0.29 & 19.16/0.24 & 17.65/0.16 & 20.46/0.33 & 19.28/0.25 & 18.63/0.19 \\
        Vanilla-TRAD & \textbf{21.09/0.41} & \underline{19.71}/0.28 & \textbf{19.54/0.31} & \textbf{22.56/0.47} & \textbf{20.79/0.35} & \underline{20.09/0.34} \\
        Accelerated-TRAD & \underline{20.33/0.33} & \textbf{20.68/0.36} & \underline{19.05/0.28} & \underline{21.54/0.36} & \underline{20.45/0.32} & \textbf{20.49/0.35} \\
        \bottomrule
    \end{tabular}}
\end{table}

Table \ref{tab:tab2} displays the PSNR and SSIM of the reconstructed results using different algorithms at SNRs of 30, 20, and 10 for sampling ratios of 1.6 and 1.8. It is evident that the proposed algorithms are more robust than others. The PSNR of reconstructed results in Vanilla-TRAD or Accelerated-TRAD is at least 1dB higher than that in the other three algorithms.

\subsection{Why combine TV with DD?}
We construct two numerical experiments to explore the reasons for combining TV with DD. (1) To analyze the effectiveness of combining explicit and implicit regularization, we compare the proposed algorithms with related algorithms that have no regularization, TV regularization only, and DD regularization only. (2) To determine the necessity of combining TV and DD, we replace TV with explicit sparse or low-rank regularizer, and DD with SIREN \cite{Sitzmann2020} or DIP \cite{Ulyanov2017}. We then compare these alternatives to Accelerated-TRAD. All compared algorithms are executed under the ADMM framework. The average PSNR of six reconstructed images in the experimental data is calculated.

\begin{figure}[htbp]
    \centering
    \includegraphics[width = 7cm]{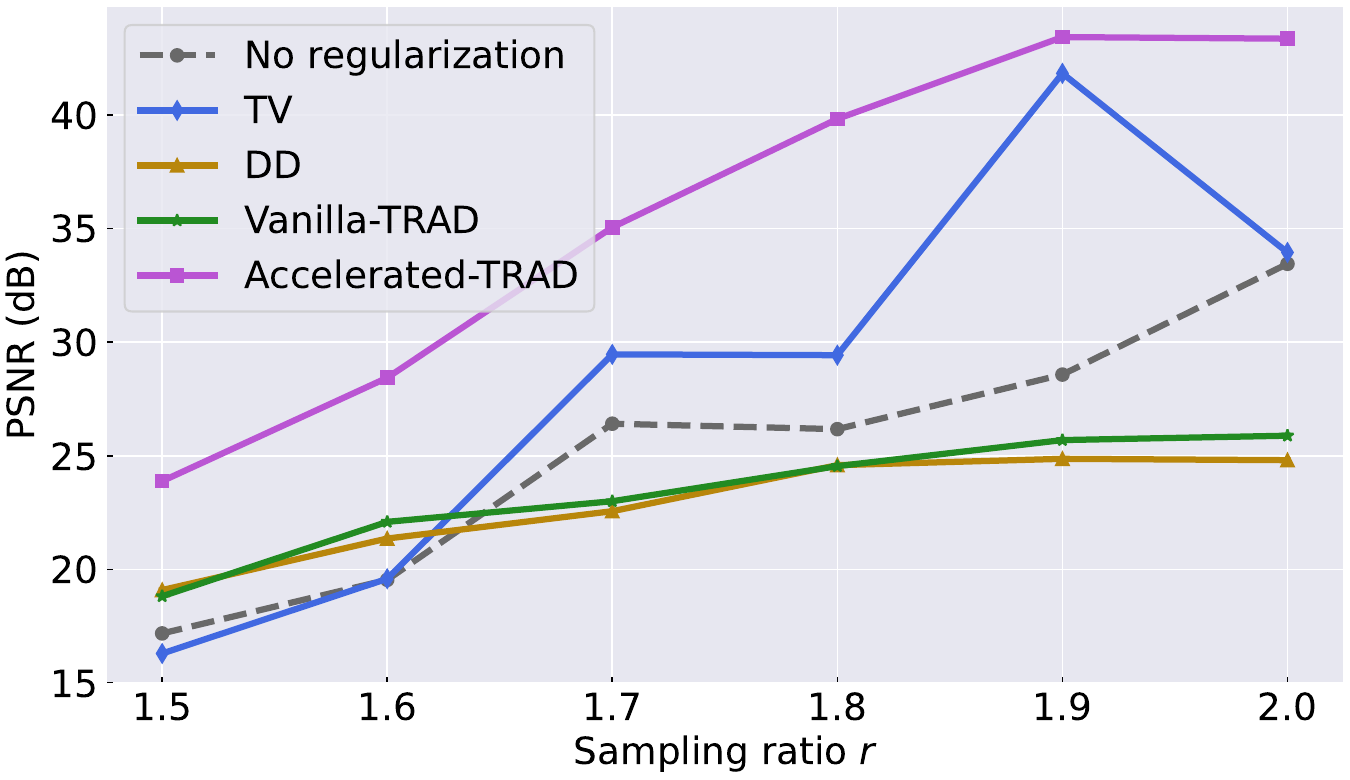}
    \caption{Quantitative results of our algorithms as well as related algorithms with no regularization, TV regularization only, and DD regularization only.}
    \label{fig:abla_psnr}
\end{figure}

\begin{figure}[htbp]
    \centering
    \includegraphics[width = 6cm]{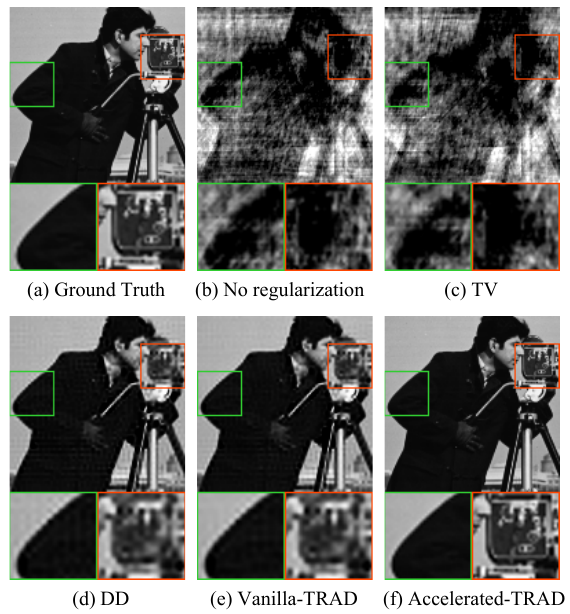}
    \caption{Visual comparison of our algorithms as well as related algorithms with no regularization, TV regularization only, and DD regularization only at the sampling ratio of 1.7.}
    \label{fig:img_ATRAD}
\end{figure}

In the first experiment, the quantitative results of five algorithms at six sampling ratios are shown in Fig. \ref{fig:abla_psnr}. A visual comparison of the reconstructed \emph{Cameraman} at the sampling ratio of 1.7 can be seen in Fig. \ref{fig:img_ATRAD}. Accelerated-TRAD performs the best and accurately recovers both high- and low-frequency information. While explicit TV regularization is effective for FPR at high sampling ratios, it may cause artifacts. In contrast, implicit DD regularization is good for FPR at low sampling ratios but leads to oversmoothed patterns. Vanilla-TRAD, which combines DD and TV regularization, can slightly alleviate this issue and improve the PSNR of reconstructed results. The accelerated technique used in Accelerated-TRAD not only reduces the time cost but also effectively plays the role of the two priors.

In the second experiment, the TV regularization is replaced with sparse regularization by $l_1$ norm $\Vert \xx \Vert_1$ and low-rank regularization by nuclear norm $\Vert \xx \Vert_\ast$, respectively. Additionally, the DD is replaced with SIREN and DIP, respectively. The corresponding network structures are described as follows. The learning rate and number of iterations are consistent with Accelerated-TRAD.

\begin{figure}
    \centering
    \includegraphics[width = 7cm]{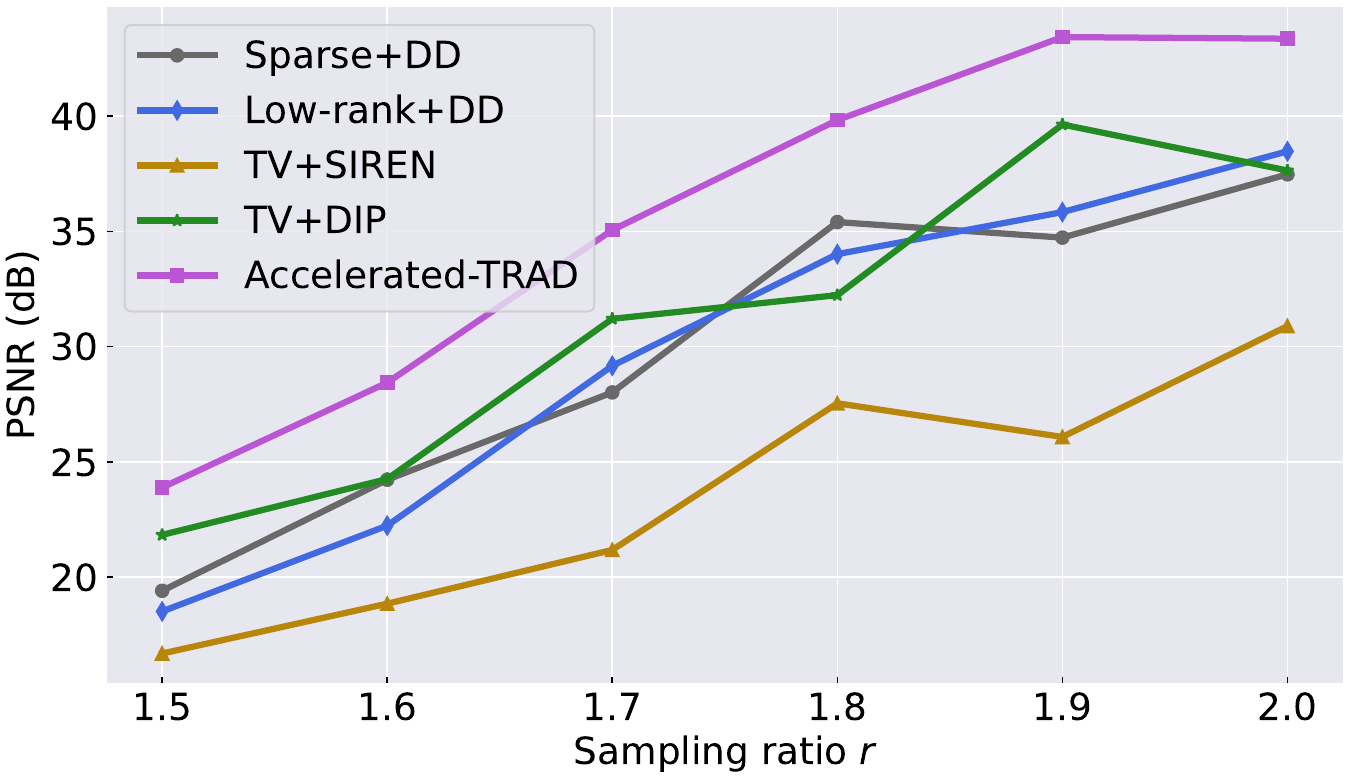}
    \caption{Quantitative results of Accelerated-TRAD and its alternatives using other explicit and implicit regularization combinations. The number of parameters in DD, SIREN, and DIP are 108160, 50049, and 1037193, respectively.}
    \label{fig:other_reg}
\end{figure}

\begin{itemize}
    \item SIREN \cite{Sitzmann2020}: An implicitly defined network is utilized to represent continuous signals and their derivatives. The network is composed of multilayer perceptrons (MLPs) with periodic sinusoidal activation. It consists of four hidden layers with 128 channels, takes image coordinates as input, and outputs pixel values.
    \item DIP \cite{Ulyanov2017}: An untrained generative network uses a U-Net-like "hourglass" architecture with skip connections. It includes upsampling and downsampling layers with channel numbers and kernel sizes of \{16, 32, 64, 128, 128\} and \{3, 3, 3, 3, 3\}, respectively. Upsampling and downsampling are implemented using nearest neighbor interpolation and convolution strides. Skip connections have channel numbers and kernel sizes of \{4, 4, 4, 4, 4\} and \{1, 1, 1, 1, 1\}, respectively. The network input follows a uniform distribution of $\mathcal{U}(0, 0.1)$.
\end{itemize}

Fig. \ref{fig:other_reg} shows the quantitative results of Accelerated-TRAD and its alternatives using different regularization combinations. It is apparent that Accelerated-TRAD, which combines TV and DD, outperforms the other regularization combinations. Additionally, the number of parameters in DD is only one-tenth that in DIP, and slightly higher than that in SIREN.

While experimental results have shown the reasons for combining TV and DD, the underlying mechanism behind this combination has yet to be theoretically explained. This is worth further research.

\subsection{The impact of parameters} \label{impact}
\begin{figure}[htbp]
    \centering
    \includegraphics[width = 7cm]{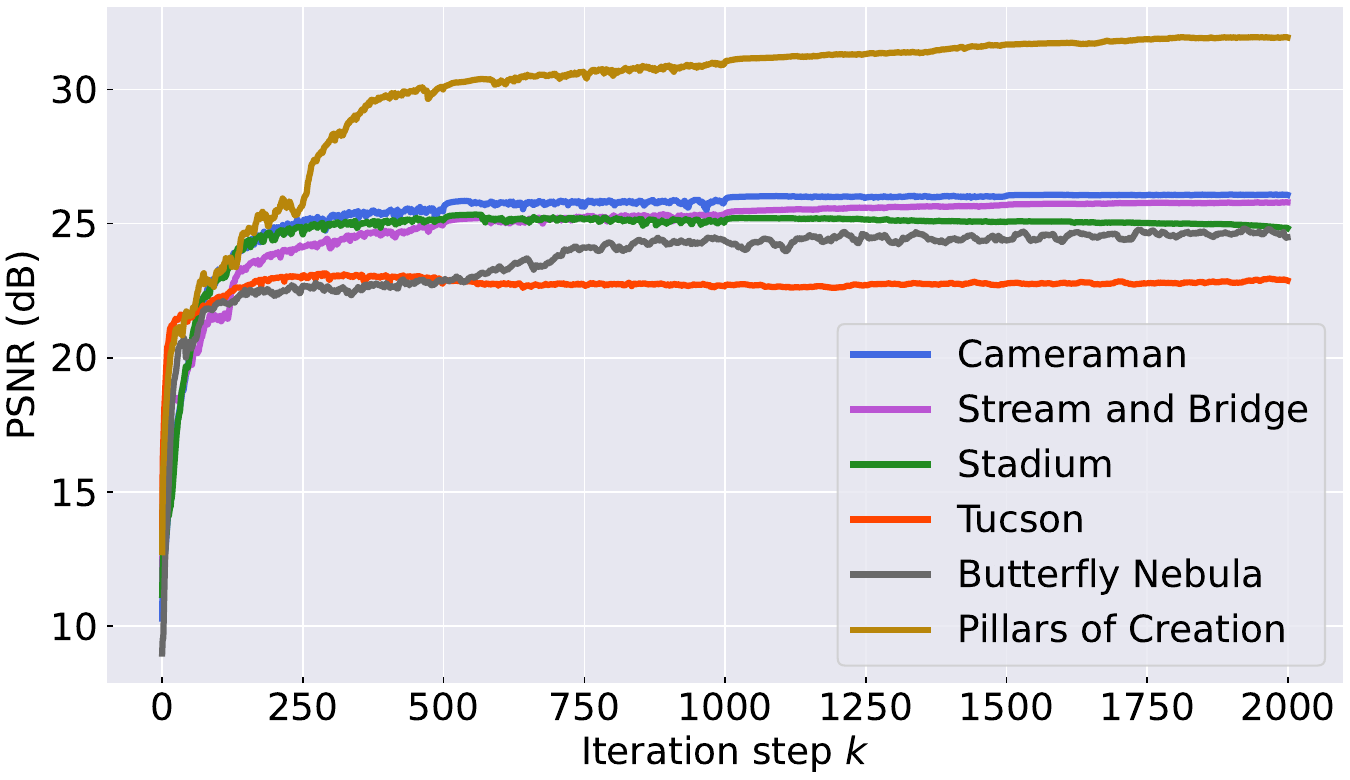}
    \caption{PSNR of reconstructed results during the iterations of Vanilla-TRAD at the sampling ratio of 2.0.}
    \label{fig:dis_psnr}
\end{figure}

This subsection examines the impact of the parameters $\lambda$ and $\kappa_3$ in Accelerated-TRAD. Firstly, we explore the influence of $\kappa_3$, which represents the iteration step at which the weight $\mu_k$ begins to decrease. We use Vanilla-TRAD to perform FPR on the experimental data, due to the consistency of Vanilla-TRAD and Accelerated-TRAD in the first $\kappa_3$ iterations. Fig. \ref{fig:dis_psnr} shows the PSNR of the reconstructed images during the iterations. We observe that the PSNR of almost all images stabilizes when $k>500$. This indicates that Accelerated-TRAD is not sensitive to $\kappa_3$, as long as $\kappa_3 > 500$. Furthermore, $\kappa_3$ generally does not need to change with different $\xx$. 

\begin{figure}[htbp]
    \centering
    \includegraphics[width = 7cm]{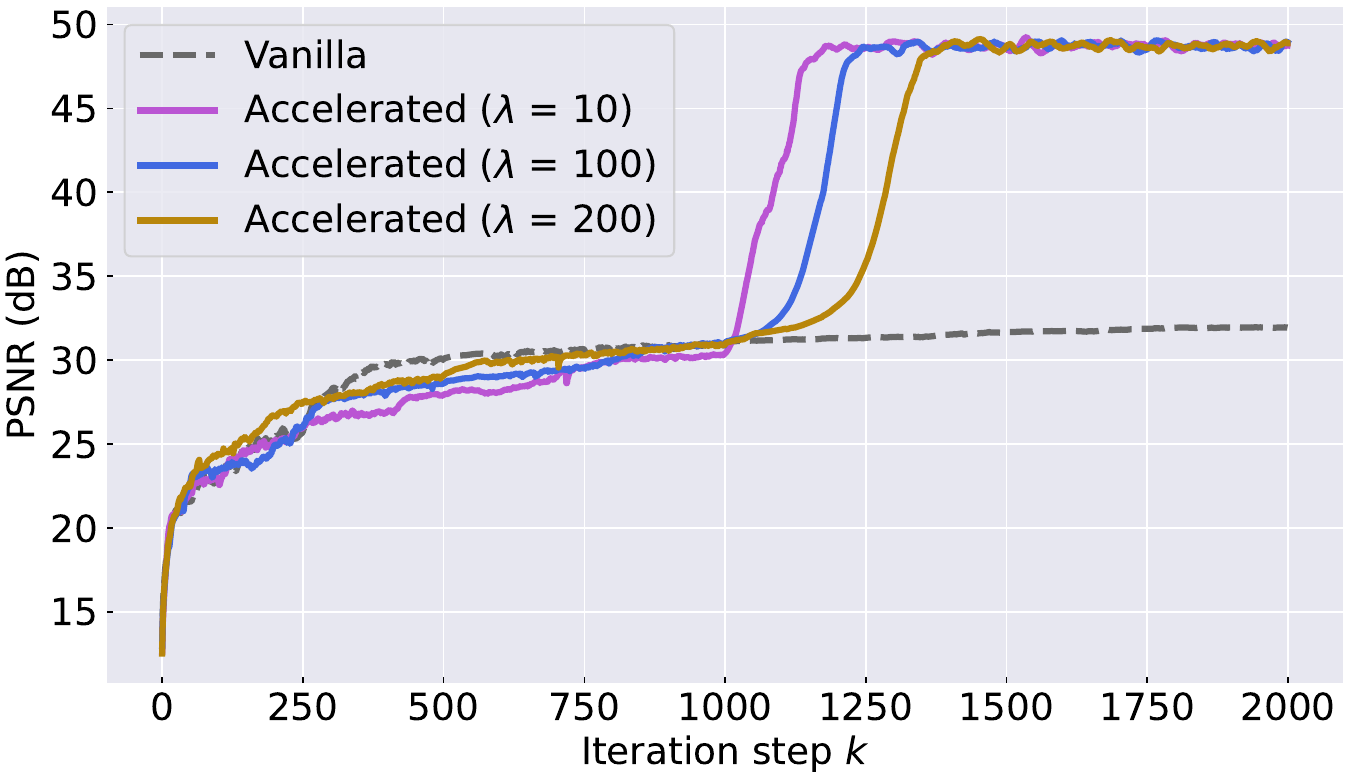}
    \caption{Comparison of Vanilla-TRAD and Accelerated-TRAD with different $\lambda$. The algorithms perform FPR on \emph{Pillars of Creation} at the sampling ratio of 2.0. The computational times for Vanilla-TRAD and Accelerated-TRAD with $\lambda = 10$, $\lambda = 100$ and $\lambda = 200$ are 70.77s, 35.37s, 41.22s and 47.67s, respectively.}
    \label{fig:lambda_psnr}
\end{figure}

Next, we consider the impact of $\lambda$, which determines the decrease rate of $\mu_k$. When $\kappa_3 = 1000$, we apply Accelerated-TRAD to \emph{Pillars of Creation} at $\lambda$ values of 10, 100, and 200, respectively, and compare with Vanilla-TRAD. The PSNR of the reconstructed images during the iterations is shown in Fig. \ref{fig:lambda_psnr}. It can be observed that Accelerated-TRAD is superior to Vanilla-TRAD in terms of both reconstructed quality and computational time. In Accelerated-TRAD, the PSNR of reconstructed results increases at different rates when $\lambda$ takes different values, but eventually reaches similar results. A small $\lambda$ tends to result in a low computational time, due to the rapid decrease of the weight $\mu_k$.

\section{Conclusion}
This paper proposes an untrained NN-based algorithm called Vanilla-TRAD that effectively combines explicit TV regularization and implicit untrained generative prior with the ADMM framework, along with its accelerated version, Accelerated-TRAD. The untrained generative prior is beneficial for FPR with few measurements, while TV regularization is beneficial for recovering high-frequency information in the image. Numerical experiments confirm the effectiveness and necessity of combining the two priors. The acceleration technique in Accelerated-TRAD reduces the computational cost and enhances the role of the two priors in improving the reconstructed quality. Extensive experiments demonstrate that Accelerated-TRAD achieves fast and high-quality reconstruction of FPR under various settings. This paper presents a new approach to FPR under limited measurements and computational resources. In the future, we plan to explore a general paradigm that combines the prior based on image gradient with the untrained generative prior.

\bibliographystyle{elsarticle-num}
\bibliography{main}

\end{document}